\documentclass[11pt]{article}

\pdfoutput=1

\usepackage{times}
\usepackage{bm}
\usepackage{graphicx}
\usepackage{amsfonts}
\usepackage{amssymb}
\usepackage{amsmath}
\usepackage{graphicx}
\usepackage{natbib}
\usepackage{color}
\usepackage{graphicx}
\usepackage{fullpage}
\usepackage{setspace}
\usepackage{multirow}
\usepackage{array}
\usepackage{float}

\newcommand{\by}{\mathbf{y}}
\newcommand{\bY}{\mathbf{Y}}

\newcommand{\V}{p}
\newcommand{\R}{\mathbb{R}}
\newcommand{\G}{\mathcal{G}}
\newcommand{\D}{\mathcal{D}}
\newcommand{\SG}{\Sigma_{\mathcal{G}}}

\usepackage{geometry}
\floatstyle{ruled}
\newfloat{algorithm}{H}{lof}
\floatname{algorithm}{Algorithm}

\begin{document}

\title{An empirical Bayes procedure for the selection of Gaussian graphical models}
\author{Sophie Donnet \\ CEREMADE \\ Universit\'e Paris Dauphine, France \and 
Jean-Michel Marin\footnote{Corresponding author: jean-michel.marin@univ-montp2.fr} \\ 
Institut de Math\'ematiques et Mod\'elisation de Montpellier \\ Universit\'e Montpellier 2, France} 

% \institute{Sophie Donnet \at CEREMADE \\ Universit\'e Paris Dauphine \\ \email{donnet@ceremade.dauphine.fr}  
% \and Jean-Michel Marin \at Institut de Math\'ematiques et Mod\'elisation de Montpellier \\ Universit\'e Montpellier 2 \\
% place Eug\`ene Bataillon \\  Case Courrier 051 \\ 34095 Montpellier cedex 5 \\ \email{jean-michel.marin@univ-montp2.fr}
% }

% \footnote{Corresponding author:}
% \footnote{\textsc{jean-michel.marin@univ-montp2.fr}} \\ Institut de Math\'ematiques et Mod\'elisation de Montpellier \\ Universit\'e Montpellier 2}

\date{}

\maketitle

\begin{abstract}
A new methodology for model determination in decomposable graphical Gaussian models 
\citep{dawid:lauritzen:1993} is developed. The Bayesian paradigm  is used and, for each given graph,
a hyper inverse Wishart prior distribution on the covariance matrix is considered. This prior distribution
depends on hyper-parameters. \\ It is well-known that the models's posterior distribution is sensitive to the
specification of these hyper-para\-meters and no completely satisfactory method is registered.
In order to avoid this problem, we suggest adopting an empirical Bayes strategy, that is a strategy for which
the values of the hyper-parameters are determined using the data. Typically, the hyper-parameters are fixed to
their maximum likelihood estimations. In order to calculate these maximum likelihood estimations, we suggest 
a Markov chain Monte Carlo version of the Stochastic Approximation EM algorithm. \\ Moreover, we introduce a new sampling scheme
in the space of graphs that improves the \textit{add and delete} proposal of \cite{armstrong:etal:2009}.
We illustrate the efficiency of this new scheme on simulated and real datasets.

% \keywords{Gaussian graphical models \and decomposable models \and empirical Bayes \and
% Stochastic Approximation EM \and Markov Chain Monte Carlo}

\vspace{0.5cm}\textbf{Keywords:} Gaussian graphical models, decomposable models, empirical Bayes, Stochastic Approximation EM,
Markov Chain Monte Carlo

\end{abstract}

\newpage

\section{Gaussian graphical models in a Bayesian Context}\label{introduction}
Statistical applications in genetics, sociology, biology , etc often lead to complicated interaction
patterns between variables. Graphical models have proved to be powerful tools to  represent the conditional
independence structure of a multivariate distribution : the nodes represent the variables and the absence
of an edge between two vertices indicates some conditional independence between the associated variables.  

Our paper presents a new approach for estimating the graph structure in Gaussian graphical model. A very large literature
deals with this issue in the Bayesian paradigm:
\cite{dawid:lauritzen:1993,madigan:raftery:1994,giudici:green:1999,jones:etal:2005,armstrong:etal:2009,carvalho:scott:2009}.
For a frequentist point of view, one can see \cite{drton:perlman:2004}.

We suggest here an empirical Bayes approach: the parameter of the prior are estimated from the data.
Parametric empirical Bayes methods have a long history, with major
developments evolving in the sequence of papers by \cite{efron:morris:1971,efron:morris:1972a,efron:morris:1972b,
efron:morris:1973a,efron:morris:1973b,efron:morris:1976a,efron:morris:1976b}.
Empirical Bayes estimation falls outside the Bayesian para\-digm. However, it has proven to be an effective technique
of constructing estimators that performs well under both Bayesian and frequentist criteria.
Moreover, in the case of decomposable Gaussian graphical models, it gives a default and objective way for constructing
prior distribution. The theory and applications of empirical Bayes methods are given by \cite{morris:1983}.

In this Section, we first recall some results on Gaussian graphical models, then we justify the use of the empirical
Bayes strategy.

\subsection{Background on Gaussian graphical models}

Let $\mathcal{G}=(V,E)$ be an undirected graph with vertices $V=\{1,\ldots,p\}$ and set of edges \\$E=$ $\{e_1,\ldots,e_t\}$,
($\forall i=1,\ldots,t$, $e_i \in V\times V$). 
Using the notations of  \cite{giudici:green:1999}, we first recall the definition of a decomposable graph. 
A graph or subgraph is said to be complete if all pairs of its vertices are joined by edges.
Moreover, a complete subgraph that is not contained within another complete subgraph is called a clique.
Let $\mathcal{C}=\{C_1,\ldots,C_k\}$ be the set of the cliques of  an undirected
graph. \\ An order of  the cliques $(C_1,\ldots,C_k)$ is said to be perfect if $\forall i=2,\ldots,k$, $\exists  h=h(i)\in\{1,\ldots,i-1\}$
such that $S_i=C_i\cap\cup_{j=1}^{i-1}C_i\subseteq  C_h$.
$\mathcal{S}=\{S_2,\ldots,S_k\}$ is the set of separators associated to the perfect order $\{C_1,\ldots,C_k\}$.
An undirected graph admitting  a perfect order  is said to be decomposable. Let $\D_p$ denote  the set of decomposable graphs with
$p$ vertices. For more details, one can refer to \cite{dawid:lauritzen:1993}, \cite{lauritzen:1996} (Chapters 2, 3 and 5)
or \cite{giudici:green:1999}.

\begin{figure}[htb]
\begin{center}
\includegraphics[width=\columnwidth]{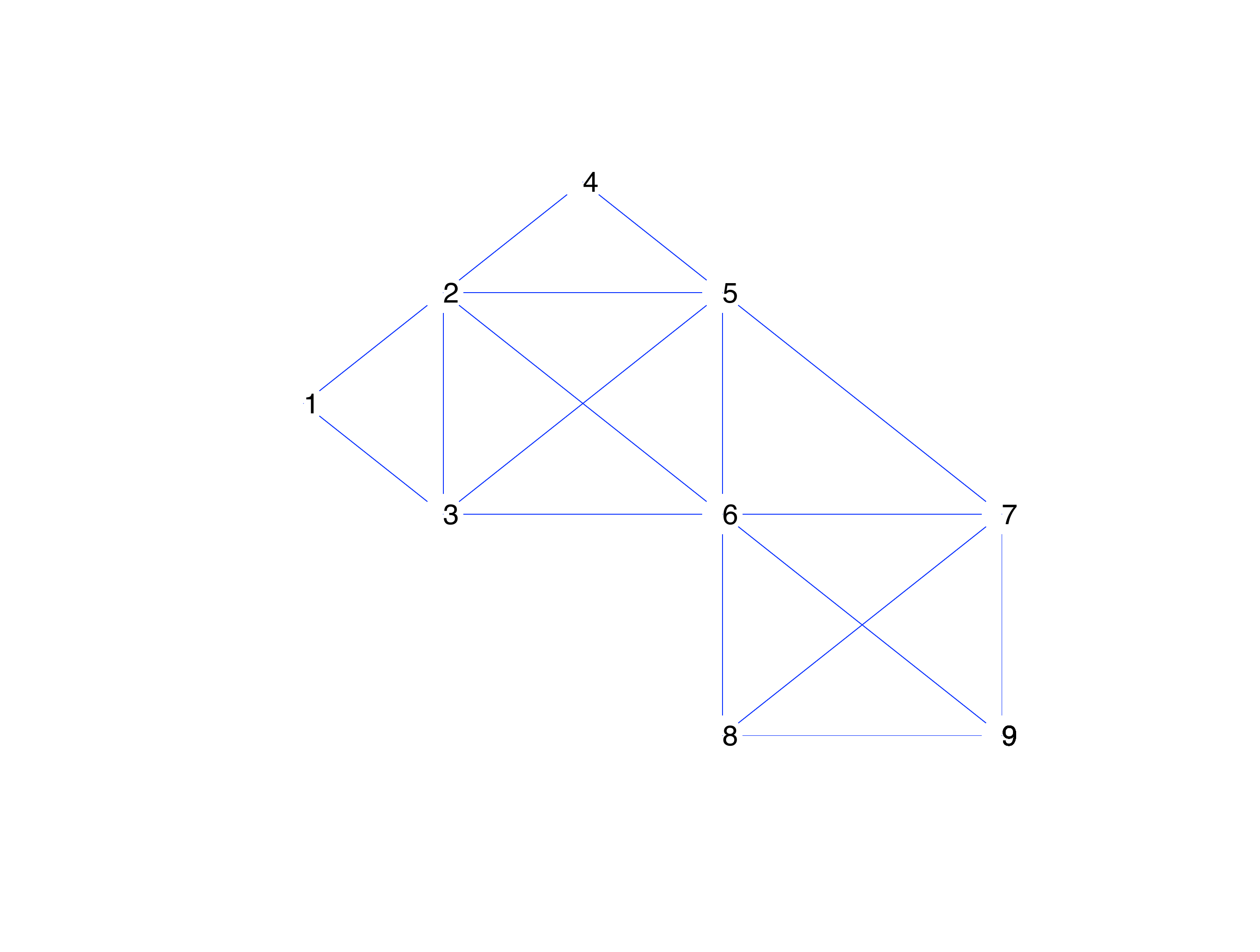}
\end{center}
\caption{Example of decomposable graph}
\label{graphe_simule}
\end{figure}

% \begin{example} 
The graph drawn in Figure \ref{graphe_simule} -- and used as benchmark in numerical Section \ref{simu}-- 
is decomposable. Indeed,  the set of cliques   $C_1=\{1,2,3\}$, $C_2=\{2,3,5,6\}$, $C_3=\{2,4,5\}$, $C_4=\{5,6,7\}$ and $C_5=\{6,7,8,9\}$ with
associated separators  $S_2=\{2,3\}$, $S_3=\{2,5\}$, $S_4=\{5,6\}$ and $S_5=\{6,7\}$ forms a perfect order. 
% \end{example}

% \begin{remark} 
Note that, with $p$ vertices, the total number of possible graphs is $2^{p(p-1)/2}$,  $p(p-1)/2$ being 
the number of possible edges.  The total number of decomposable graphs with $p$ vertices can be calculated for moderate values of $p$. For instance,
if $p=6$, among the $32\,768$ possible graphs, $18\,154$ are decomposable (around $55\%$); if $p=8$, then  $30\,888\,596$ of the $268\,435\,456$ possible
graphs are decomposable (around $12\%$). 
% \end{remark}

A pair $(A,B)$ of subsets of the vertex set $V$ of an undirected graph $\mathcal{G}$ is said to form
a decomposition of $\mathcal{G}$ if  (1) $V=A\cup B$ ,  (2) $A\cap B$ is complete and  (3) $A\cap B$ separates $A$ from $B$ 
ie any path from a vertex in $A$ to a vertex in $B$ goes through $A\cap B$.

To each vertex $v\in V$, we associate a random variable $y_v$. 
For $A\subseteq V$, $\by_A$ denotes  the collection of random variables
$\{y_v:v\in A\}$. To simplify the notation,  we set  $\by=\mathbf{y}_V$.
The probability distribution of $\by$ is said to be Markov with respect to $\mathcal{G}$,
if for any decomposition $(A,B)$ of $\mathcal{G}$, $\by_A$ is independent of $\by_B$ given
$\by_{A\cap B}$. A graphical model is a family of distributions on $\by$ verifying the  Markov property  with respect to a graph. 

A Gaussian graphical model, also called covariance
selection model (see \cite{dempster:1972}), is such that
\begin{equation}
\mathbf{y}|\G,\SG\sim
\mathcal{N}_p\left(\boldsymbol{\mu},\SG\right)\,,
\label{eqn:model}
\end{equation}
where $\mathcal{N}_p\left(\boldsymbol{\mu},\SG\right)$ denotes the $p$-variate Gaussian distribution
with expectation $\boldsymbol{\mu}\in\mathbb{R}^p$ and $p\times p$ symmetric definite positive covariance matrix $\SG$.
$\SG$ has to ensure the Markov property with respect to $\G$.
In the Gaussian case, $\mathbf{y}$ is Markov with respect to $\mathcal{G}=(V,E)$ if and only if 
$$
(i,j)\notin E\Longleftrightarrow \left(\Sigma_\mathcal{G}^{-1}\right)_{(i,j)}=0\,,
$$
where $A^{-1}$ denotes the inverse of the matrix $A$. $\SG^{-1}$ is called the concentration matrix. 

In the following, we suppose that we observe a sample $\bY=(\by^1,\ldots,\by^n)$ from model (\ref{eqn:model}) with mean parameter
$\boldsymbol{\mu}$ set to zero. The data are expressed as a deviation from the sample mean. This centering strategy is standard in the literature,
however the technique developed here can be easily  extended to the case $\boldsymbol{\mu}\neq \mathbf{0}_p$.

The density of $\bY$ is a function of multivariate Gaussian densities on the cliques and separators of $\G$.
More precisely, let $\mathcal{C}$ and $\mathcal{S}$ denote respectively the sets of the cliques and separators of $\G$ corresponding to a perfect order for $\G$.  
We have : 
\begin{equation}
f(\bY|\SG,\G)= \prod_{i=1}^n \left\{ \frac{\prod_{C \in \mathcal{C}} \phi_{|C|}\left(\by^i_C|(\SG)_C\right)}
{\prod_{S \in \mathcal{S}}\phi_{|S|}\left(\by^i_S|(\SG)_S\right)}\right\}\,,
\end{equation}
where for every subset of vertices $A$,  $|A|$ denotes its cardinal and $(\SG)_A$ is the restriction of $(\SG)$ to $A$ i.e.
$\left\{(\SG)_{i,j}\right\}_{i \in A, j \in A}$ and $\by_A=(\by_j)_{j \in A}$. $\phi_q\left(\cdot|\Delta\right)$ is the
$q$-variate Gaussian density with mean $\mathbf{0}_q$ and $q\times q$ symmetric definite positive covariance matrix
$\Delta$. 

From a Bayesian perspective, we are interested in the posterior probabilities 
\begin{equation}
\pi(\G|\bY) \propto\pi(\G) \int f(\bY|\SG,\G) \pi(\SG | \G) d\SG\,,
\end{equation} 
for specific priors $\pi(\SG|\G)$ and $\pi(\G)$. In the following, we discuss the choice of these
prior distributions.

\subsection{Prior distributions specification}\label{MGG}

\paragraph{Prior and posterior distributions for the covariance matrix}

$\;$ \\

Conditionally on $\G$, we set an Hyper-Inverse Wishart (HIW) distribution as prior distribution on $\Sigma_\mathcal{G}$:
$$
\Sigma_\mathcal{G}|\mathcal{G},\delta,\Phi\sim \mbox{HIW}_\mathcal{G}\left(\delta,\Phi\right)\,,
$$
where $\delta>0$ is the degree of freedom  and $\Phi$ is a $p\times p$ symmetric positive definite location matrix.
This distribution is the unique hyper-Markov distribution such that, for every clique $C \in \mathcal{C}$, $(\SG)_C \sim IW(\delta, \Phi_C)$ 
with density
{\small \begin{equation}\label{eq:densIW}
\begin{array}{ccl}
\pi\left((\SG)_C | \delta,\Phi_C\right)&=& h^{IW}_{\G_C}(\delta,\Phi_C)\left[\det (\SG)_C\right]^{-\frac{\delta + 2 |C|}{2}}\\
 && \exp\left\{-\frac{1}{2} \text{tr} \left[(\SG)_C^{-1} \Phi_C\right]\right\}\,,\\
\end{array}
\end{equation}}
where $h^{IW}_{\G_C}(\delta,\Phi_C)$ is the normalizing constant: 
\begin{equation}\label{eq:constIW} 
h^{IW}_{\G_C}(\delta,\Phi_C)= \frac{\det \left(\frac{\Phi_C}{2}\right) ^{(|C|+\delta-1)/2}}{\Gamma_{|C|}\left(\frac{|C|+\delta-1}{2}\right)}\,,
\end{equation}
where $\det(\cdot)$ and $\text{tr}(\cdot)$ are respectively the determinant and trace and 
$\Gamma_v$ is the multivariate  $\Gamma$-function with parameter $v$:
$$
\Gamma_v(a) = \pi^{v(v-1)/4} \prod_{j=1}^v \Gamma[a+(1-j)/2]\,.
$$

The full joint density is:
\begin{equation}\label{eq:densHIW}
\pi(\SG|\G,\delta,\phi) = \frac{\prod_{C \in \mathcal{C}} \pi\left((\SG)_C|\delta,\Phi_C\right)}{\prod_{S \in \mathcal{S}}\pi\left((\SG)_S |\delta,\Phi_S\right)}\,.
\end{equation}

Conditionally on $\mathcal{G}$, the HIW distribution is conjugate. 
The posterior distribution of $\Sigma_\mathcal{G}$ is given by \citep{giudici:1996}:
\begin{equation}\label{eqn:posterior}
\Sigma_\mathcal{G}|\bY,\mathcal{G},\delta,\Phi\sim
\mbox{HIW}\left(\delta+n,\Phi+S_{\bY} \right)\,.
\end{equation}
where $S_{\bY}= \sum_{i=1}^n\by^i \: \mathstrut^t\by^i$, $\mathstrut^t v$ denoting the transpose of $v$.

Moreover for such a prior distribution, the marginal likelihood for any graph $\G$ is a simple function of the HIW prior and posterior normalizing
constants $h_{\G}(\delta, \Phi)$ and $h_{\G}(\delta+n, \Phi + S_{\bY})$ \citep{giudici:1996}:
\begin{equation}\label{eq:margdensity}
f(\bY|\mathcal{G},\delta,\Phi)= \frac{h_{\G}( \delta, \Phi)}{ (2\pi)^{-np/2} h_{\G}( \delta+n, \Phi + S_{\bY})}\,.
\end{equation}
where   $h_{\G}(\delta, \Phi)$ is the normalizing constant of the HIW distribution which can be computed explicitly in decomposable graphs from the normalizing constants of the inverse Wishart cliques
and separators densities (\ref{eq:densIW}-\ref{eq:constIW}-\ref{eq:densHIW}) : 
$$ 
\begin{array}{ccc}
h_{\G}( \delta, \Phi) &=&\frac{ \prod_{C \in \mathcal{C}}h^{IW}_{\G_C}(\delta,\Phi_C)}{\prod_{S \in \mathcal{S}} h^{IW}_{\G_S}(\delta,\Phi_S)}\,.
\end{array}
$$

% \begin{remark}
Note that \cite{roverato:2002} extends the Hyper-Inverse Wishart distribution to non- decomposable cases. 
Moreover, a general treatment of priors for decomposable models is given by \cite{letac:massam:2007}.
% \end{remark}

\paragraph{Prior and posterior distributions for the graphs}

$\;$ \\

The prior distribution in the space of decomposable graphs has been widely discussed in the literature. 
The naive choice is to use the standard uniform prior distribution: 
\begin{equation*}
\pi(\mathcal{G})\propto 1\,.
\label{eq:priorg}
\end{equation*}
One great advantage of this choice is simplifying the calculus but it can be criticized. Indeed,
with $p$ vertices, the number of possible edges  is equal to $m = \frac{p(p-1)}{2}$ and, in the case of a uniform prior
over all graphs, the prior number of edges has its mode around $m/2$ which is typically too large.

An alternative to this prior is to set a Bernouilli distribution of parameter $r$ on the inclusion or not of each edge \citep{jones:etal:2005,carvalho:scott:2009}
\begin{equation}
\pi(\mathcal{G}|r)\propto r^{k_\mathcal{G}} (1-r)^{m-k_\mathcal{G}}\,,
\label{eq:priorg-scott}
\end{equation}
where $k_\G$ is the number of edges of $\G$.
The parameter $r$ has to be calibrate. If $r = 1/2$, this prior resumes to the uniform one. 

In the following we consider this prior distribution and give an empirical estimation of $r$.

Using (\ref{eq:margdensity}) and (\ref{eq:priorg-scott}), we deduce easily that the density of the posterior
distribution in the space of decomposable graphs satisfies:
\begin{equation}
\pi\left(\G|\bY,\delta,r,\Phi\right)\propto \frac{h_{\G}( \delta, \Phi)}{h_{\G}( \delta+n, \Phi + S_{\bY})} \pi(\G|r)\,.
\label{eqn:postgraph}
\end{equation}

This posterior distribution is known to be sensitive to the specification of the hyper-parameters
$r$, $\delta$ and $\Phi$ (see \cite{jones:etal:2005,armstrong:etal:2009}). To tackle this problem various
strategies have been developed. In the following, we supply a short review of these methods and offer 
an alternative one. 

\paragraph{Choice of the hyper-parameters $\delta$, $r$ and $\Phi$}

$\;$ \\

In a fully Bayesian context, as proposed by \cite{giudici:green:1999}, a hierarchical prior modelling can be used.
In this approach,  $\delta$and $\Phi$ are considered as random quantities and a prior distribution is assigned to 
those parameters ($r$ is fixed to $1/2$). This strategy does not completely solve the problem since the prior distributions
on $\delta$ and $\Phi$ also depend on hyper-parameters which are difficult to calibrate.

An other strategy consists in fixing the values of $\delta$, $r$ and $\Phi$ as in \cite{jones:etal:2005}. In that paper,  $r$ is set to $\frac{1}{p-1}$ encouraging sparse graphs.  
They choose $\delta=3$ which is the minimal integer
such that the first moment of the prior distribution on $\SG$ exists. Finally,  they set  $\Phi=\tau I_p$ and using the fact that the mode of the marginal
prior for each variance  terms $\sigma_{ii}$ is equal to $\tau/(\delta+2)$, $\tau$ is fixed to $\delta+2$  if the data set is standardized.

An intermediate strategy is suggested by \cite{armstrong:etal:2009}.  First, they fix  the value of $\delta$ to $4$ \footnote{In fact, they set
$\delta=5$ but they consider that $\boldsymbol{\mu}$ is unknown with uniform prior distribution: this situation corresponds to the case
$\delta=4$ when $\boldsymbol{\mu}=\mathbf{0}_p$.}  assessing that such a value gives a suitably non-informative prior for $\SG$.
Then, they consider different possibilities for $\Phi$, all of the form $\Phi=\tau A$ where the matrix $A$ is fixed.
In all cases, for the hyper-parameter $\tau$, they use a uniform prior distribution on the interval $[0,\Gamma]$ where $\Gamma$ is very large. 
Finally, they also use a hierarchical prior on $r$ : $r \sim \beta(1,1)$, which  leads to
$$\pi(G) \propto \left( \begin{array}{c} m\\ k_{\G}\end{array} \right) ^{-1}$$ 
by integration.  $ \left( \begin{array}{c} m\\ k_{\G}\end{array} \right) $ is  the binomial coefficient.   

This hierarchical prior of $r$ is also used in \cite{carvalho:scott:2009}. In that paper, they suggest a 
HIW $g$-prior approach with $g=1/n$. This approach consists of fixing
$\delta=1$ and $\Phi=S_\bY/n$.

In our point of view, $\delta$ measures the amount of information in the prior relative to the sample (see (\ref{eqn:posterior})): 
we suggest setting  $\delta$ to $1$ such that the prior weight is the same as the weight of one observation.
As pointed out by \cite{jones:etal:2005}, for this particular choice, the first moment of the prior distribution on $\SG$ does not exist.
However, for $\delta=1$, the prior distribution is proper and we fail to see any argument in favour of the existence of
a first moment.

The structure of $\Phi$ can be discussed and various forms exist in the literature (see \cite{armstrong:etal:2009} for instance).
In this paper,  we  standardise the data and use $\Phi=\tau I_p$. This choice leads to sparse graph: on average
each variable has major interactions with a relatively small number of other variables. In that context, $\tau$ plays the role of a shrinkage
factor and has to be carefully chosen on the appropriate scale. 

In this paper, we recommend to use an empirical Bayes strategy and to fix $(\tau,r)$ to its maximum
likelihood estimation for which computation is a challenging issue. To tackle this point, a Markov Chain Monte Carlo (MCMC) version
of the Stochastic Approximation EM (SAEM) algorithm is used. 

The SAEM algorithm is presented in Section \ref{SAEM}. In Section \ref{MCMC}, a new Metropolis-Hasting algorithm is introduced.
Then, the proposed methodology is tested on real and simulated datasets.

\section{An empirical Bayes procedure via the SAEM-MCMC algorithm}\label{SAEM}

In the following, we set $\theta = (\tau, r) \in \R^{*+} \times ]0,1[$.  
In order to compute the maximum likelihood estimation of $\theta$, we need to optimize in $\theta$ the following function
\begin{equation}\label{eq:marg-lik}
f(\bY|\theta)\propto \sum_{\G\in\mathcal{D}_p}\left\{\frac{h_\G(\delta,\tau I_p)}{h_\G(n+\delta,\tau I_p+S_\bY)}\right\}\pi(\G| r) \,
\end{equation}
If the number of vertices is greater than $10$, the number of decomposable graphs is so huge that it is not possible
to calculate the sum over $\mathcal{D}_p$. In that case, we consider the use of  the Expectation-Maximization (EM) algorithm
developed by \\ \cite{dempster:laird:rubin:1977},  noting the fact that the data  $\bY=$ \\ $(\by^1,\ldots,\by^n)$ are issued from
the partial observations of the complete data $(\bY,\G,\SG)$. However, for such a data augmentation scheme, the E-step of the EM algorithm is not explicit and
we have to resort to a stochastic version  of the EM algorithm, like:
\begin{enumerate}
\item the S-EM scheme introduced by \cite{celeux:diebolt:1992} and \cite{diebolt:celeux:1993} where the E-step is replaced by
a single simulation from the distribution of $(\G,\SG)$ given $\bY$ and $\theta$;
\item the MC-EM or the MCMC-EM algorithms \\ where the E-step is replaced by some Monte Carlo approximations \\ \citep{mclachlan:krishnan:2008};
\item the SAEM algorithm introduced by \cite{delyon:lavielle:moulines:1999} where the E-step is divided into a simulation step and a stochastic
approximation step;
\item the SAEM-MCMC algorithm \\ \citep{kuhn:lavielle:2004} which extends the SAEM scheme, the ``exact'' simulation step being replaced by a simulation
from an ergodic Markov chain.
\end{enumerate}
The S-EM, MC-EM and SAEM methods require to simulate a realization
from the distribution of $(\G,\SG)$ given $\bY$ and $\theta$. We are not able to produce a realization exactly distributed from the distribution
of $(\G,\SG)$ given $\bY$ and $\theta$. We use the SAEM-MCMC algorithm which
just requires some realizations from an ergodic Markov chain with stationary distribution $(\G,\SG)|\bY,\theta$.
In a first part, we recall the EM algorithm principles and present the SAEM-MCMC scheme.
In a second part, we detail its application to Gaussian graphical models and prove its convergence. 

\subsection{The Stochastic Approximation version of the EM algorithm}\label{em-algo}

The EM algorithm is competitive when the maximization of the function 
$$
\theta \rightarrow Q(\theta|\theta')=\mathbb{E}_{\SG,\G|\bY,\theta'} \left\{\log f(\bY,\SG,\G|\theta)\right\}
$$
is easier than the direct maximization of the mar\-ginal likelihood (\ref{eq:marg-lik}).  
The EM algorithm is a two steps iterative procedure.  More precisely, 
at the $k$-th iteration, the E-step consists  of evaluating 
$Q_{k}(\theta)=Q(\theta \, \vert \,\widehat{\theta}_{k-1})$ while the M-step updates
$\widehat{\theta}_{k-1}$ by maximizing $Q_{k}(\theta)$.  

For complicated models where the E-step is untractable, \cite{delyon:lavielle:moulines:1999}
introduce the Sto\-chastic Approximation EM algorithm (SAEM) replacing
the E-step by a stochastic approximation of $Q_{k}(\theta)$. At iteration $k$, the E-step is 
divided into a simulation step (S-step) of  $\left(\SG^{(k)},\G^{(k)}\right)$ with the posterior distribution 
$(\SG,\G)| \bY,\widehat{\theta}_{k-1}$ and a stochastic approximation step (SA-step):
$$
\begin{array}{cll}
Q_{k}(\theta) &=&(1-\gamma_k) Q_{k-1}(\theta)+ \\
&&\gamma_k \log f(\bY,\SG^{(k)},\G^{(k)}|\widehat{\theta}_{k-1})
\end{array}
$$
where $(\gamma_k)_{k\in \mathbb{N}}$ is a sequence of positive numbers decreasing to
zero. When the joint distribution of $(\bY,\SG,\G)$ belongs to the exponential family, the
SA-step reduces to the stochastic approximation on the minimal exhaustive statistics. 
The M-step remains the same. One of the benefits of the SAEM algorithm is the low-level dependence
on the initialization $\theta_0$, due to the stochastic approximation of the SA-step. 

In Gaussian graphical models, we cannot generate directly a realization from the conditional distribution of
$(\SG,\G)$ given $\bY$ and $\widehat{\theta}_{k-1}$.
For such cases, \cite{kuhn:lavielle:2004} suggest to replace the simulation step by a MCMC scheme which consists of
generating $M$ realizations from an ergodic Markov chain with stationary distribution
$\SG,\G|\bY,\widehat{\theta}_{k-1}$ and use the last simulation in the SAEM algorithm.
\cite{kuhn:lavielle:2004} prove the convergence of the estimates sequence provided by this SAEM-MCMC algorithm
towards a maximum of the function $f(\bY|\theta)$ under general conditions for the exponential family.

\subsection{The SAEM-MCMC algorithm on Gaussian graphical models }\label{SAEM-appli}

\noindent In this section, we detail the application of the SAEM-MCMC algorithm to the Gaussian graphical model
introduced in Section \ref{MGG}. More precisely, we give the expression of the complete log-likelihood and
of the minimal sufficient statistics. \cite{lavielle:lebarbier:2001} applied the same methodology on a change-point problem.

The complete log-likelihood $f(\bY,\G,\SG|\theta)$ can be decomposed into three terms: 
 \begin{multline}\label{complete_log_likelihood}
\log f(\bY,\G,\SG|\theta)=\log f(\bY|\G,\SG)\\
+ \log \pi(\SG|\G,\tau)
+\log \pi(\G|r)\,.
\end{multline}

On the right-hand side of equation (\ref{complete_log_likelihood}), the first quantity is independent of $\theta$ thus, it will not take part
in its estimation.  Using the fact that we only consider decomposable graphs and the definition of the Hyper Inverse Wishart distribution, the second term of the right-hand side of Equation (\ref{complete_log_likelihood}) can be developed :

{\small
\begin{minipage}{0.98 \columnwidth}
\begin{multline*}
\log \pi (\SG|\G,\tau)  = \sum_{C \in \mathcal{C}} \frac{|C|(|C|+\delta-1)}{2} \log(\tau) \\
-\log\Gamma_{|C|}\left( \frac{|C|+\delta-1}{2}\right)
-\frac{\delta+2|C|}{2}\log \det (\SG)_C\\  
-\sum_{S \in \mathcal{S}} \left[ \frac{|S|(|S|+\delta-1)}{2} \log(\tau)  -\log \Gamma_{|S|}\left( \frac{|S|+\delta-1}{2}\right)\right. \\
\left. -  \frac{\delta+2|S|}{2} \log \det (\SG)_S \right] - \frac{\tau}{2}  \text{tr}(\SG^{-1})\,.
\end{multline*}
\end{minipage}}

Furthermore, 
$$
\log \pi(\G | r) = k_{\G} \log\left(\frac{r}{1-r}\right) +   m \log(1-r)\,.
$$

As a consequence, there exists   $\Psi$  a function of $(\bY,\SG,\G,\delta)$
independent of $\theta=(\tau,r)$ such that 
{\small \begin{multline}
\label{logvrais}
 \log f(\bY,\G,\SG|\tau) =  \Psi\left(\bY,\SG,\G,\delta\right) \\
   +\frac{\delta-1}{2}p \log(\tau) + m \log(1-r)+   \frac{1}{2}\times\\ \left< \left( 
   \begin{array}{c}
   \sum_{C\in \mathcal{C}}|C|^2-\sum_{S \in \mathcal{S}}|S|^2 \\
   \text{tr}(\SG^{-1})\\
   k_{\G} 
\end{array}
  \right) ,
  \left(
    \begin{array}{c}
  \log(\tau)\\
    -\tau\\
    \log\left(\frac{r}{1-r}\right)
\end{array}
\right)\right>\,, 
\end{multline}}

where $\left<\cdot,\cdot \right>$ is the scalar product of $\R^3$.
Finally, following (\ref{logvrais}), the complete likelihood function belongs to the exponential family and 
the minimal sufficient statistic $S=(S_1,S_2,S_3)$ is such that:
\begin{eqnarray*}
S_1(\bY,\G,\SG)&=&  \sum_{C\in \mathcal{C}}|C|^2-\sum_{S \in \mathcal{S}}|S|^2 \\
S_2(\bY,\G,\SG)&=&\text{tr}(\SG^{-1})\\
S_3(\bY,\G,\SG)&=&k_{\G} \,.
\end{eqnarray*}
In an exponential model, the SA-step of the SAEM-MCMC algorithm reduces to the approximation of
the minimal sufficient statistics. Thus, we can now write the three steps of the  SAEM-MCMC algorithm:
let $\left(\gamma_k\right)_{k \in \mathbb{N}}$ be a sequence of  positive numbers such that
$\sum_k \gamma_k=\infty$  and  $\sum_k \gamma_k^{2}<\infty$.

\newpage

\rule[-2.5ex]{\columnwidth}{0.3mm}
\textbf{Algorithm 1} SAEM-MCMC algorithm\\
\rule[3.1ex]{\columnwidth}{0.3mm}
\vspace{-3em}

\begin{itemize}
\item[(1)] Initialize $ \widehat{\theta}^{(0)}$, $s_1^{(0)}$,  $s_2^{(0)}$ and $s_3^{(0)}$.
\item[(2)] At iteration $k$, \\
\end{itemize}
\vspace{-1em}
$\bullet$  [S-Step] generate  $\G^{(k)},\SG^{(k)}$ from $M$ iterations of a MCMC procedure -- detailed in Section 
\ref{MCMC} -- with $\G,\SG |\bY,\widehat{\theta}^{(k-1)}$ as stationnary distribution.\\
$\bullet$ [SA-Step] update $\left(s_i^{(k)}\right)_{i=1,2,3}$ using a stochastic approximation scheme: $i=1,2,3$
$$
s_i^{(k)}=  s_i^{(k-1)} + \gamma_k \left(S_i(\bY,\G^{(k)},\SG^{(k)}) - s_i^{(k-1)}\right)\,.
$$
$\bullet$ [M-Step] maximize the joint log-likelihood (\ref{logvrais}):  
$$
\widehat{\tau}^{(k)}= \frac{(\delta-1)p+s_1^{(k)}}{s_2^{(k)}}\quad
\widehat{r}^{(k)}=  \frac{s_3^{(k)}}{m}\,.
$$
\begin{itemize}
\item[(3)] Set $k=k+1$ and return to (2) until convergence.
\end{itemize}
\rule[3.1ex]{\columnwidth}{0.3mm}

The convergence of the estimates sequence supplied by this SAEM-MCMC algorithm  is ensured by the results of \cite{kuhn:lavielle:2004}.
Indeed, first, the complete likelihood   belongs to the exponential family and  the regularity assumptions required by \cite{kuhn:lavielle:2004}
(assumptions \textbf{M1-M5} and \textbf{SAEM2}) are easily verified.  
Secondly, the convergence requires the ergodicity of the Markov Chain generated at S-step towards the stationary distribution
that is the distribution of $\G,\SG|\bY,\widehat{\theta}^{(k-1)}$.
Finally, the properties of  $(\gamma_k)_{k \in \mathbb{N}}$  allow to apply
the results of \cite{kuhn:lavielle:2004} and we conclude   that the estimates  sequence $(\widehat{\theta}^{(k)})_{k \in \mathbb{N}}$
converges almost surely towards a (local) maximum of the function $f(\bY|\theta)$. 
 
\section{A new Metropolis-Hastings sampler}\label{MCMC}

At each iteration $k$ of the SAEM algorithm, a couple $(\G,\SG)$ has to be generated under the posterior distribution
$\G,\SG|\bY,\theta^{(k-1)}$. As described in \cite{giudici:green:1999}, \cite{brooks:giudici:roberts:2003} and \cite{wong:carter:kohn:2003},
this simulation can be achieved using a variable dimension MCMC scheme like the reversible jump algorithm.
In case of an HIW prior distribution on $\SG$, the marginal likelihood is available
in closed form (\ref{eq:margdensity}) and, therefore, there is no need to resort to a variable dimension MCMC scheme. 

At iteration $k$ of the SAEM algorithm, the simulation of $(\G,\SG)^{(k)}$ can be achieved through the following
two steps procedure:
\begin{itemize}
\item[$\bullet$] [S1-step]  
$\G^{(k)}\sim \pi(\G|\bY,\theta^{(k-1)})$
\item[$\bullet$] [S2-step] $ \SG^{(k)}\sim \pi(\SG|\G^{(k)},\bY,\theta^{(k-1)})$
\end{itemize}
According to (\ref{eqn:posterior}), the second step [S2-step]  of this procedure resolves into the simulation of HIW distributions the
principle of which is detailed in \cite{carvalho:massam:west:2007}. 

For the first step [S1-step], we have to resort to an MCMC algorithm
but not of variable dimension since the chain is generated in the decomposable graphs space with $p$ vertices.

To sample for the posterior in the space of graphs, \cite{armstrong:etal:2009} use the fact that the marginal likelihood is available in closed form
and introduce a Metropolis-Hastings (MH) algorithm. At iteration $t$, their \textit{add and delete} MH proposal consists of 
picking uniformly at random an edge such that the current graph with or without this edge stays decomposable;
and deducing the proposed graph by deleting the generated edge to the current graph if it contains this edge or adding
the generated edge otherwise.  

Let $\G$ be the current graph, $G^{-}_{\G}$ the set of decomposable graphs derived from $\G$ by removing an edge
and  $G^{+}_{\G}$ the set of decomposable graphs derived from $\G$ by adding an edge. 
For pedagogical reasons, we present here an \textit{add and delete} MH sampler slightly different from the one of
\cite{armstrong:etal:2009}. In our proposal, we first decide at random if we try to delete or to add an edge.
The two schemes has exactly the same properties. Our \textit{add and delete} algorithm is initialized on $\G^{(0)}$
and the following procedure is repeated until the convergence is reached.

\rule[-2.5ex]{\columnwidth}{0.3mm}
\textbf{Algorithm 2} \textit{Add and Delete} MH proposal \\
\rule[3.1ex]{\columnwidth}{0.3mm}
\vspace{-2em}

At iteration $t$, 
\begin{itemize}
\item[(a)] Choose at random (with probability $1/2$) to delete or add an edge to $\G^{(t-1)}$.
\item[(a.1)] If delete an edge, enumerate $G^{-}_{\G^{(t-1)}}$ and generate $\mathcal{G}^p$ according to the uniform distribution on $G^{-}_{\G^{(t-1)}}$.
\item[(a.2)] If add an edge, enumerate $G^{+}_{\G^{(t-1)}}$ and generate $\G^p$ according to the uniform distribution on  $G^{+}_{\G^{(t-1)}}$ .
\item[(b)] Calculate the MH acceptance probability \\$\rho(\G^{(t-1)},\G^p)$ such that $\pi(\G|\bY,\theta)$ is
the invariant distribution of the Markov chain.
\item[(c)] With probability $\rho(\G^{(t-1)},\G^p)$, accept $\G^p$ and set $\G^{(t)}=\G^p$,
otherwise reject $\G^p$ and set $\G^{(t)}=\G^{(t-1)}$.
\end{itemize}
\rule[3.1ex]{\columnwidth}{0.3mm}

The acceptance probability $\rho(\G^{(t-1)},\G^p)$ is equal to $\alpha(\G^{(t-1)},\G^p)\wedge1$ where 
$$
\alpha(\G^{(t-1)},\G^p) = \frac{\pi(\G^p| \bY,\delta,r,\Phi)}{\pi(\G^{(t-1)}| \bY,\delta,r,\Phi)}\frac{q(\G^{(t-1)}| \G^p)}{q(\G^p|\G^{(t-1)})}$$
with 
$$
\frac{q(\G^{(t-1)}| \G^p)}{q(\G^p|\G^{(t-1)})} = \left\{
\begin{array}{cl} 
\frac{|G^{+}_{\G^{(t-1)}}|}{|G^{-}_{\G^p}|} & \mbox{ if  add} \\
\frac{|G^{-}_{\G^{(t-1)}}|}{|G^{+}_{\G^p}|} & \mbox{ if  delete} 
\end{array}
\right.
$$
Note that because in general $ |G^{+}_{\G^{(t-1)}}| \neq |G^{-}_{\G^p}|$,   the proposal distribution is not symmetric.
The ratio $ \frac{\pi(\G^p| \bY,\delta,r,\Phi)}{\pi(\G^{(t-1)}| \bY,\delta,r,\Phi)}$ is evaluated with  formula (\ref{eqn:postgraph}). 

The enumerations of $G^{-}_{\G^{(t-1)}}$ and $G^{+}_{\G^{(t-1)}}$ are not obvious and can be time-consuming.
To tackle this point, we apply the results of \cite{giudici:green:1999} characterizing the set of moves (\textit{add and delete})
which preserve the decomposability of the graph. These criteria lead to a fast enumeration.

\cite{armstrong:etal:2009} prove that this scheme\footnote{In \cite{armstrong:etal:2009}, the step on the space of graphs represents a Gibbs step of
an hybrid sampler (as already explained, they consider a hierarchical model where that the hyper-parameter $\tau$ is a random variable).}
is more efficient than the variable dimension sampler of \cite{brooks:giudici:roberts:2003}, which is itself an improvement of  the reversible jump algorithm
proposed by \cite{giudici:green:1999}. Their proposal is clearly irreducible and, therefore, the theoretical convergence of the produced Markov \\ Chain
towards the stationary distribution \\ $\pi(\G|\bY,\tau)$ is ensured, following standard results on MH schemes.

\vspace{0.5em}
However, in practice, the space of decomposable graphs is so large that the chain may take quite some  time to reach the invariant
distribution. To improve this point, we introduce a data-driven MH kernel which uses the informations contained in
the inverse of the empirical covariance matrix. To justify this choice, recall that,because of the Gaussian graphical model
properties, if the inverse empirical covariance between vertices $i$ and $j$ is near zero, we can presume that there is no edge
between vertices $i$ and $j$. Then, during the MH iterations, if the current graph contains an edge between vertices $i$ and $j$,
it is legitimate to propose   removing this edge. The same type of reasoning can be done if the absolute value of the inverse empirical
covariance between vertices $k$ and $l$ is large. Indeed, in that case, and if during the MH iterations the current graph
does not contain an edge between vertices $k$ and $l$, it is legitimated to propose to add this edge.
With this proposal, once the random choice to add or delete an edge has been done, the proposed graph is not chosen
uniformly within the class of decomposable graphs but according to the values of the inverse empirical covariances.

\newpage

Let $K$ denote  the inverse empirical covariance matrix: $K=\left(S_\bY/n\right)^{-1}$.  $\G^{(t-1)}\setminus (i,j)$ 
(respectively $\G^{(t-1)}\cup (i,j)$) denotes the graph $\G^{(t-1)}$ where the edge $(i,j)$ has been removed (respectively added).

The Data Driven kernel is the following one : 

\rule[-2.5ex]{\columnwidth}{0.3mm}
\textbf{Algorithm 3} Data Driven MH  proposal \\
\rule[3.1ex]{\columnwidth}{0.3mm}
\vspace{-2em}

At iteration $t$, 
\begin{itemize}
\item[(a)] Choose at random to delete or add an edge to $\G^{(t-1)}$.
\item[(a.1)] If delete an edge, enumerate $G^{-}_{\G^{(t-1)}}$ and generate $\mathcal{G}^p$ according to the distribution such that
$$
\mathbb{P}\left[\G^p= \mathcal{G}^{(t-1)}\setminus (i,j) |\mathcal{G}^{(t-1)}\right] \propto \frac{1}{|K_{i,j}|}\,.
$$
\item[(a.2)] If add an edge, enumerate $G^{+}_{\G^{(t-1)}}$ and generate $\mathcal{G}^p$ according to the distribution such that
$$
\mathbb{P}\left[\G^p= \mathcal{G}^{(t-1)}\cup (i,j) |\mathcal{G}^{(t-1)}\right] \propto |K_{i,j}|\,.
$$
\item[(b)] Calculate the MH acceptance probability \\ $\rho(\mathcal{G}^{(t-1)},\mathcal{G}^p)$ such that $\pi(\G|\bY,\tau)$ is
the invariant distribution of the Markov chain.
\item[(c)] With probability $\rho(\mathcal{G}^{(t-1)},\mathcal{G}^p)$, accept $\mathcal{G}^p$ and set $\mathcal{G}^{(t)}=\mathcal{G}^p$,
otherwise reject $\mathcal{G}^p$ and set $\mathcal{G}^{(t)}=\mathcal{G}^{(t-1)}$.
\end{itemize}
\rule[3.1ex]{\columnwidth}{0.3mm}

The algorithm is initialized on $\G^{(0)}$ and the procedure is repeated until the convergence is reached.  

Finally, in view of some numerical experiments and in order to keep the good properties in terms of exploration of the standard
MH kernel, we propose to use in practice a combination of the standard  \textit{add and delete} MH kernel  and the previously presented data-driven kernel.
This point is detailed in the next section. 

\section{Numerical experiments}

In this part, we illustrate the statistical performances of our methodology on three different data sets.
The second one is a simulated example which highlights the convergence properties of the SAEM-MCMC algorithm. 
The first and third examples appeared in \cite{whittaker:1990} and have been widely used to evaluate the statistical
performance of graphical models methodology, one can see for instance \cite{giudici:green:1999,armstrong:etal:2009}.
Through these two examples, the importance of the choice of the hyper-parameters and the efficiency
of the new MCMC sampler are underlined.

\subsection{The Fret's heads dataset  \cite{whittaker:1990} }

Fret's heads dataset contains head measurements on the first and the second adult son in a sample of $n=25$ families.
The $p=4$ variables are the head length of the first son, the head breadth of the first son, the head length of the second son and
the head breadth of the second son. 61 graphs are decomposable among the $64$ possibles graphs.

We compare three different prior distributions on $(\Sigma_\G, \G)$. 
\begin{enumerate}
\item We first consider the prior distribution suggested by \cite{jones:etal:2005} e.g. 
$$
\begin{array}{ccccccc}
\delta  &=&  3  &\mbox{ and }&  r&=&1/(\V-1) \\
\Phi &=& \tau I_p &\mbox{ with }& \tau &=& \delta + 2\,.
\end{array}
$$ 

\item In a second experiment, we  use the prior distribution proposed in  \cite{carvalho:scott:2009} i.e 
$$ 
\delta = 1 \quad \Phi =  \frac{S_{\by}}{n}\,.
$$
Furthermore, $r \sim \beta(1,1)$ resulting into 
$$
\pi(G) \propto \left( \begin{array}{c} m\\ k_{\G}\end{array} \right) ^{-1}\,.
$$ 

\item Finally, we use our prior distribution e.g, 
$$
\delta=1 \,,\quad \Phi = \tau I_p\,.
$$
and a Bernouilli prior of parameter $r$ on the edges of $\G$. \\
Using  the SAEM  algorithm described previously, we estimate
$\tau$ and $r$ to
$$
\widehat{\tau}=0.3925\,, \quad \widehat{r}=0.6052\,.
$$
\end{enumerate}
 
On this example,  there are only
$61$ decomposable graphs and so we are able to compute exactly the posterior probabilities $\left\{\pi(\G | \by),\G \mbox{decomposable}\right\}$
for every prior distribution. At that point, we are interested in comparing the posterior probabilities of the five most 
probable decomposable graphs for the three previously prior distribution.  The results are resumed in Table \ref{fretsheads_compartau}. 

The empirical Bayes estimation of $\tau$ is quite smal\-ler than the value provided by the heuristic of
\cite{jones:etal:2005}. As a consequence, the posterior probabilities of graphs are really different.
Moreover, the approach of \cite{carvalho:scott:2009} gives results not in agreement with one of the two others method.
The way the hyper-parameters $\tau$ and $r$ are considered is essential, since that drastically influences the results.

\begin{table*}
\begin{tabular}{ p{3cm}m{3.3cm} m{3.3cm} m{3.3cm}}
 \hline
Prior  & \multicolumn{3}{c}{Most probable posterior graphs and posterior probability $p(G| \by)$}\\
 \hline
 \hline
 
   \cite{jones:etal:2005} & \includegraphics[keepaspectratio=TRUE,height=2.5cm]{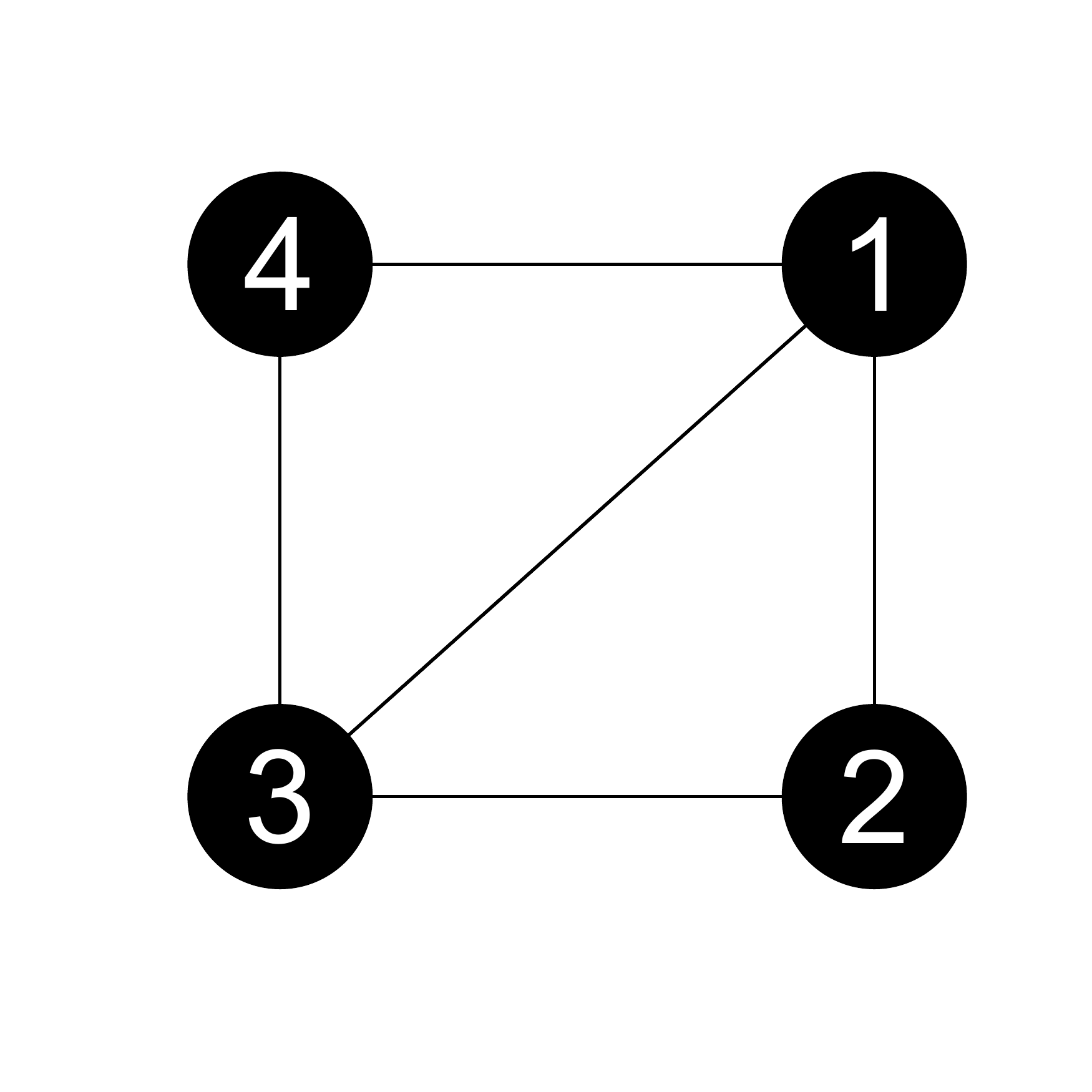} &\includegraphics[keepaspectratio=TRUE,height=2.5cm]{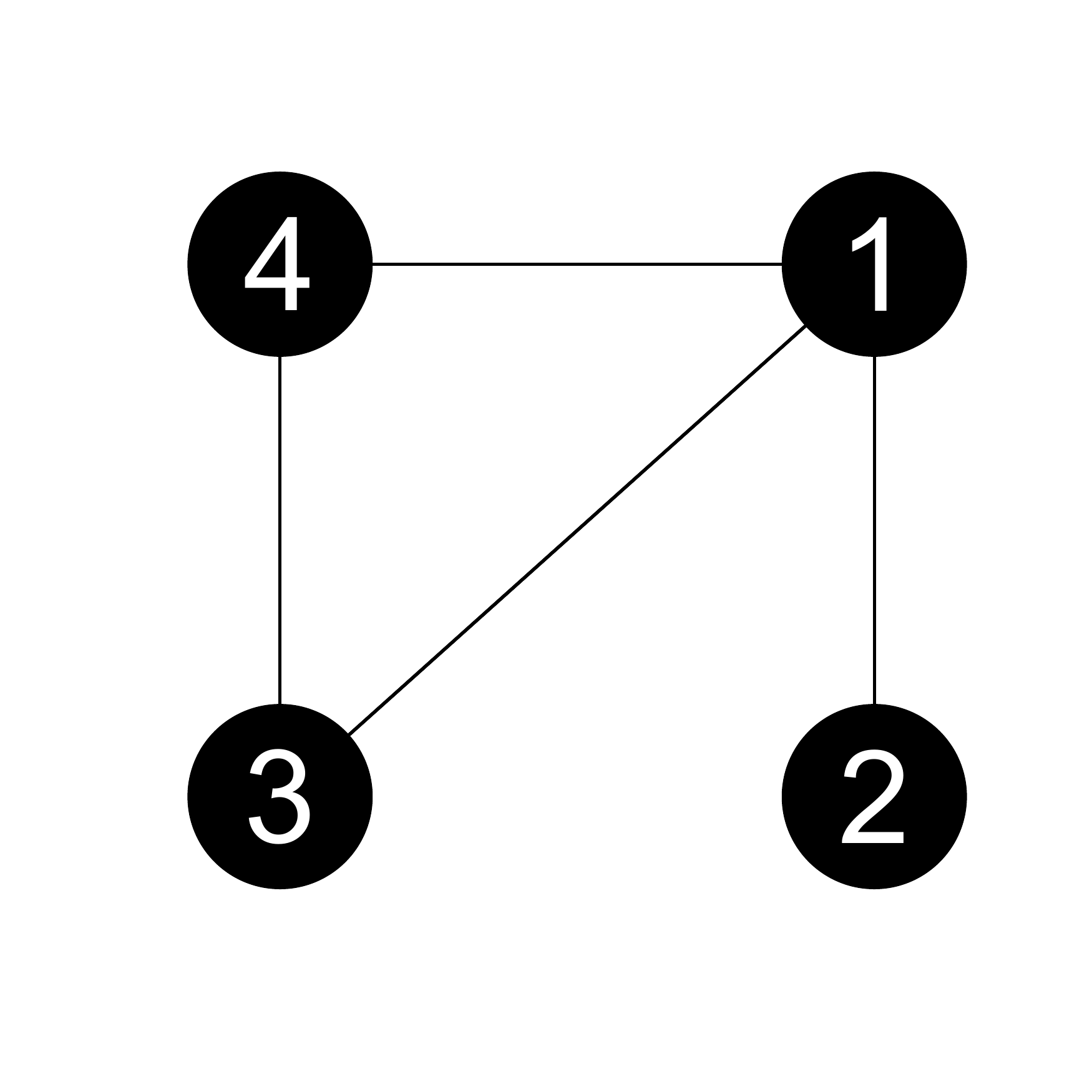}&\includegraphics[keepaspectratio=TRUE,height=2.5cm]{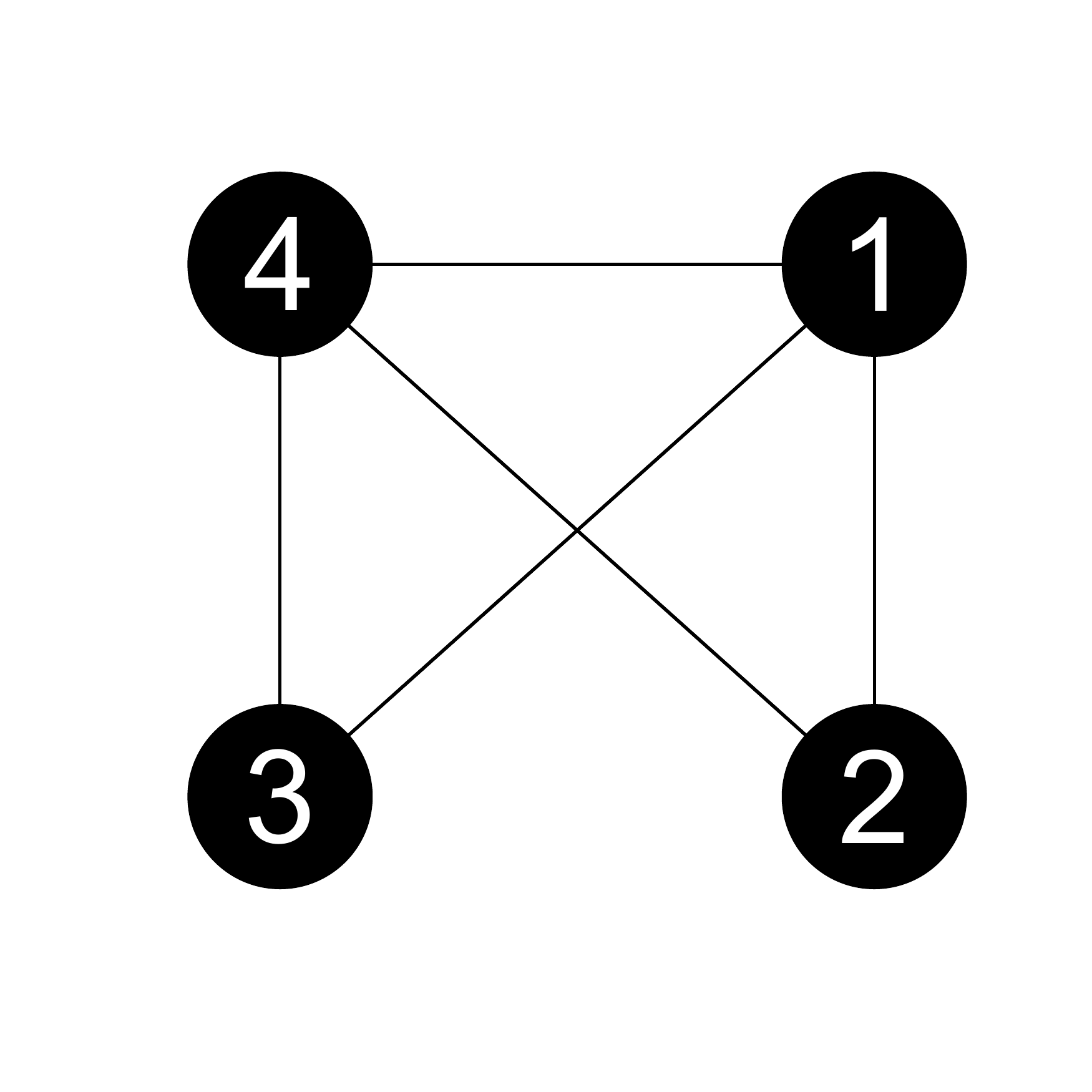}\\
  & \hspace{2em}$0.24076$ & \hspace{2em}$0.16924$ & \hspace{2em}$0.11761$\\ 

  \hline  
          \cite{carvalho:scott:2009}  &\includegraphics[keepaspectratio=TRUE,height=2.5cm]{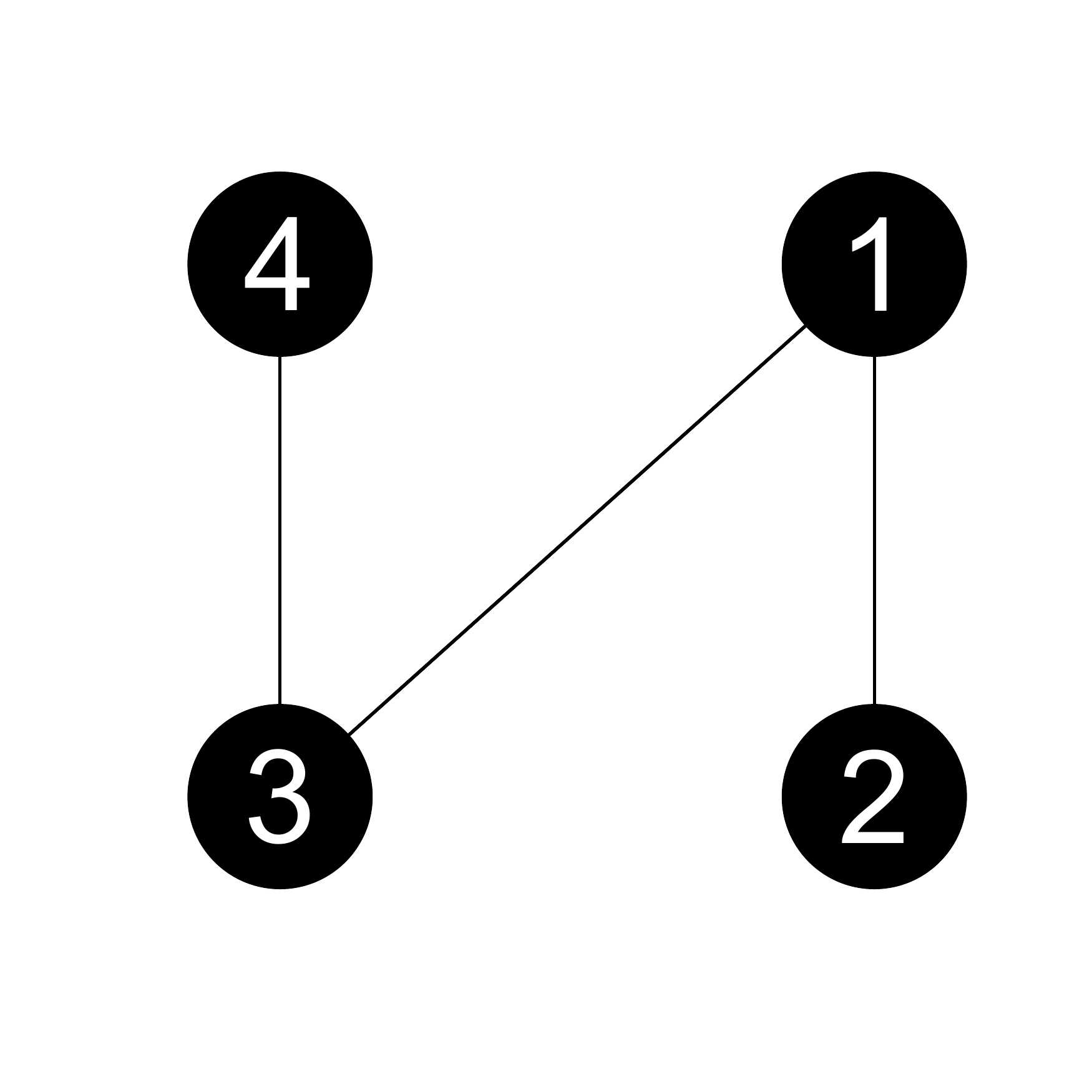} &\includegraphics[keepaspectratio=TRUE,height=2.5cm]{fret_21.pdf}&\includegraphics[keepaspectratio=TRUE,height=2.5cm]{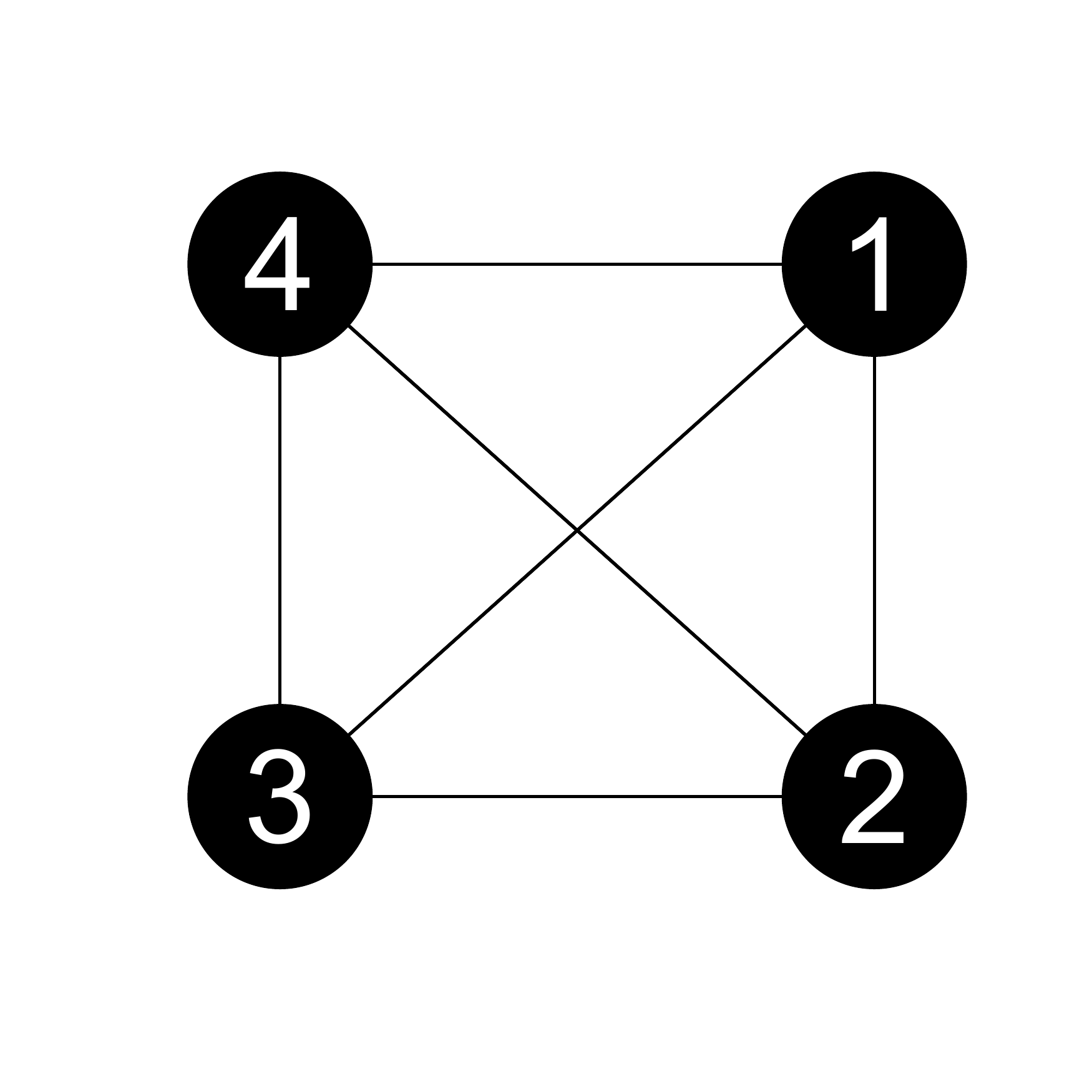}\\
       & \hspace{2em}$0.30512$ & \hspace{2em}$0.19979$ & \hspace{2em}$0.10813$\\

   \hline 
       SAEM &\includegraphics[keepaspectratio=TRUE,height=2.5cm]{fret_21.pdf} &\includegraphics[keepaspectratio=TRUE,height=2.5cm]{fret_12.pdf}&\includegraphics[keepaspectratio=TRUE,height=2.5cm]{fret_22.pdf}\\
      & \hspace{2em}$0.28613$ & \hspace{2em}$0.18219$ & \hspace{2em}$0.1264$\\
     
    \hline
       \hline
   \end{tabular}
   \caption{\textit{Fret's heads dataset} : the three most probable posterior graphs using various prior on ($\Sigma_\G, \G$). }
  \end{table*}
 
\subsection{Simulated Datasets}\label{simu}

We consider $10$ artificial datasets where $p=9$. These datasets are simulated according to model (\ref{eqn:model})
with the graph of Figure \ref{graphe_simule}.  $\tau$ , $\delta$ and $n$ are set equal to
$0.03$, $1$ and $100$ respectively. 

The SAEM-MCMC algorithm has been performed on the $10$ datasets in order to estimate the hyper-parameter $\tau$.
The algorithm is arbitrary initialized with $\hat\tau^{(0)}=0.001$ and $\hat r^{(0)}=0.5$. Given $\hat\tau^{(0)}$, $\G$ is initialized with a standard
backward procedure based on the posterior probabilities with $\hat r^{(0)}$.

The step of the stochastic approximation scheme is chosen
as recommended by \cite{kuhn:lavielle:2005}: $\gamma_k=1$ during the first iterations $1\leq k\leq K_1$, and $\gamma_k=(k-K_1)^{-1}$
during the subsequent iterations. The initial guess on $\hat\tau^{(0)}$ and $\hat r^{(0)}$ could be far from a local maximum of the likelihood function and the
first iterations with $\gamma_k=1$ allow the sequence of estimates to converge to a neighborhood of a local maximum.
Subsequently, smaller step sizes during $K-K_1$ additional iterations ensure the almost sure convergence of the algorithm
to a local maximum of the likelihood function. We implemented the SAEM-MCMC algorithm with $K_1=100$ and $K=300$. 
At the S-step of the algorithm, the Markov Chain supplied by the MCMC algorithm is of length $M=500$ during the first $5$ iterations
of the SAEM scheme and $M=10$ for the remaining iterations.
 
\begin{figure*}[htb]
\begin{center}
\includegraphics[width=0.6\textwidth]{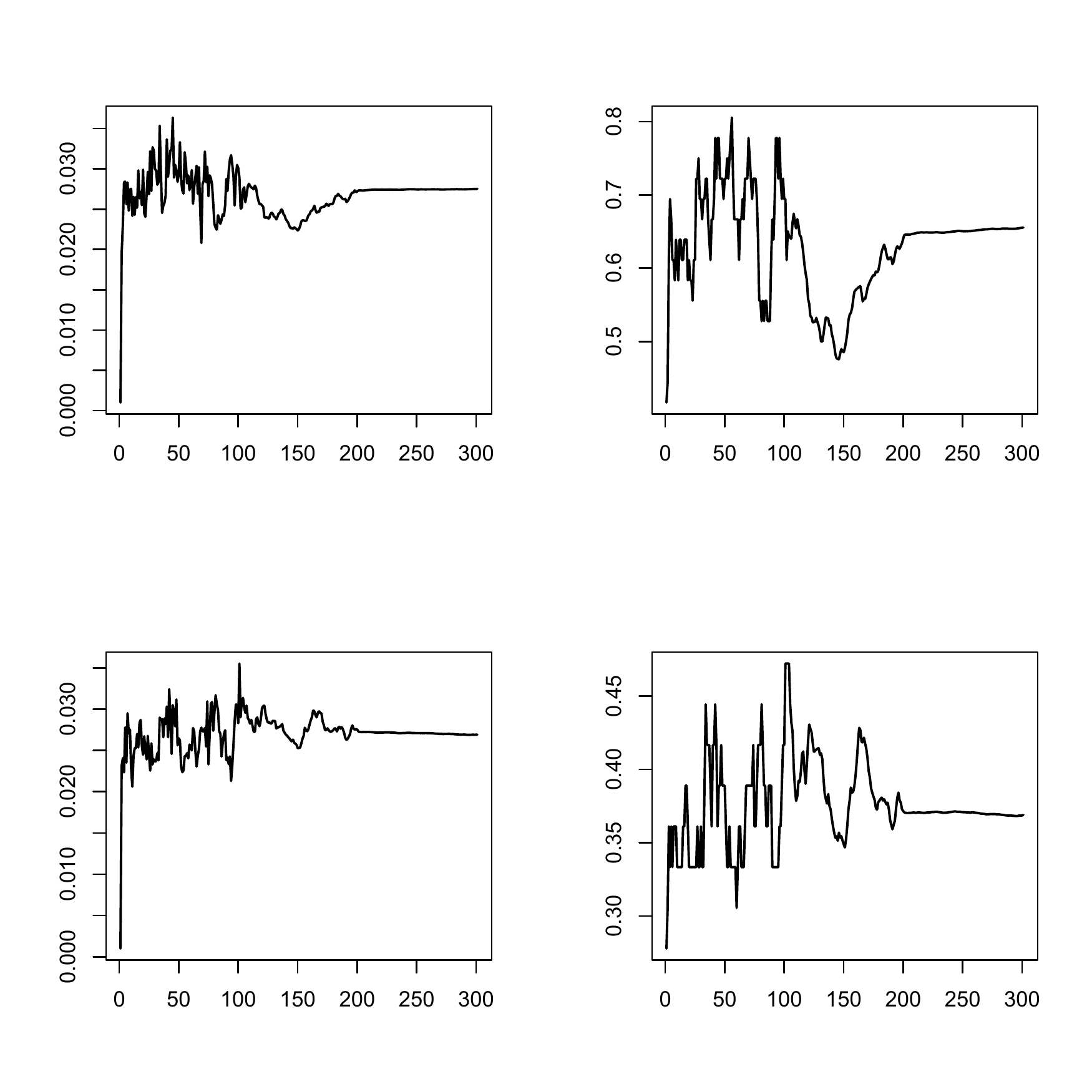} 
\end{center}
\caption{Simulated datasets: evolution of the SAEM-MCMC $\hat\tau^{(k)}$ estimations (left) and $\hat r^{(k)}$ estimations (right)
on $2$ datasets.}
\label{datasim_conv_tau}
\end{figure*}

Figure \ref{datasim_conv_tau} illustrates the convergence of the parameter estimates considering $2$ arbitrary chosen data\-sets.
The estimated sequences are represented as a function of the iteration number. During the first iterations of SAEM, the parameter
estimates fluctuate, reflecting the  Markov Chain construction. After $100$ iterations, the curves smooth but still continue to
converge towards a neighborhood of a local maximum of the likelihood function. Convergence is obtained after $300$ iterations. 
 
Considering the $10$ datasets, fir the parameter $\tau$ the relative bias is negligible and the relative root mean square error
(RMSE) amounts to $32.10\%$.  Note that the same study has been conducted with a uniform prior on $\G$
In that case, the algorithm only involves the parameter $\tau$ and the corresponding RMSE is equal to $23.5\%$.

\subsection{The Fowl bones dataset  \cite{whittaker:1990}}

This dataset concerns bone measurements which are taken from $n=276$ white leghorn fowl. The $6$ variables are
skull length, skull breadth, humerous (wings), ulna (wings), femur (legs) and tibia (legs). 
On such a dataset, the determination of the best decomposable Gaussian graphical model results
in finding the best graph within $18,154$ decomposable graphs ($55\%$ of the possible graphs).

Using this example, we aim at illustrating the fact that a careful choice of the transition kernel in the MCMC algorithm ensures
a better exploration of the support of the posterior distribution. To do this, we compare the performances of the \textit{add and delete}
proposal of \cite{armstrong:etal:2009} to those given by the data-driven one. 

In a first step, we use the SAEM-MCMC algorithm  to calibrate the value of $\tau$ and $r$. We obtain $\tau^*=0.674$ and $r^*=0.69$. 

In a second step, using this fixed value of $\tau$ and $r$, we  generate $2$ Markov chains of $110\:000$ iterations. 
The first one  is simulated using the \textit{add and delete} kernel. For the second one,
we use exclusively the \textit{add and delete} kernel  during $10\:000$ iterations : this phase of burn-in allows
a large exploration of the decomposable graphs space. During the last  $100\:000$ iterations, we alternatively  and systematically
use the \textit{add and delete} and data-driven kernels.

To illustrate the performance of this new kernel, we compute exactly the posterior probabilities \\$p(\G| \bY;\tau^*,r^*)$
for each decomposable graph  of size $\V = 6$.  We concentrate our efforts on the graphs such that  $p(\G| \bY;\tau^*,r^*) \leq 0.001$
(resulting into $107$ graphs among the $18154$ ones)  assuming the the other ones  are of small interest because nearly never reached by the Markov chains. 

For each graph of interest $\G_{int}$, we count the number of times each Markov Chain reached it (after having removed the burnin period). We finally obtain an estimation of the posterior probability by each chain:
\begin{eqnarray*}
\widehat{\pi}_1 (\G_{int} |   \bY;\tau^*,r^*) &=& \frac{|\{t ; \G_1^{(t)}=\G_{int}  \} |}{100\:000}\\
\widehat{\pi}_2 (\G_{int} |   \bY;\tau^*,r^*) &= &\frac{|\{t ; \G_2^{(t)}=\G_{int}  \} |}{100\:000}
\end{eqnarray*}
 
 \begin{figure*}[htb]
\begin{center}
\includegraphics[width=0.6\textwidth]{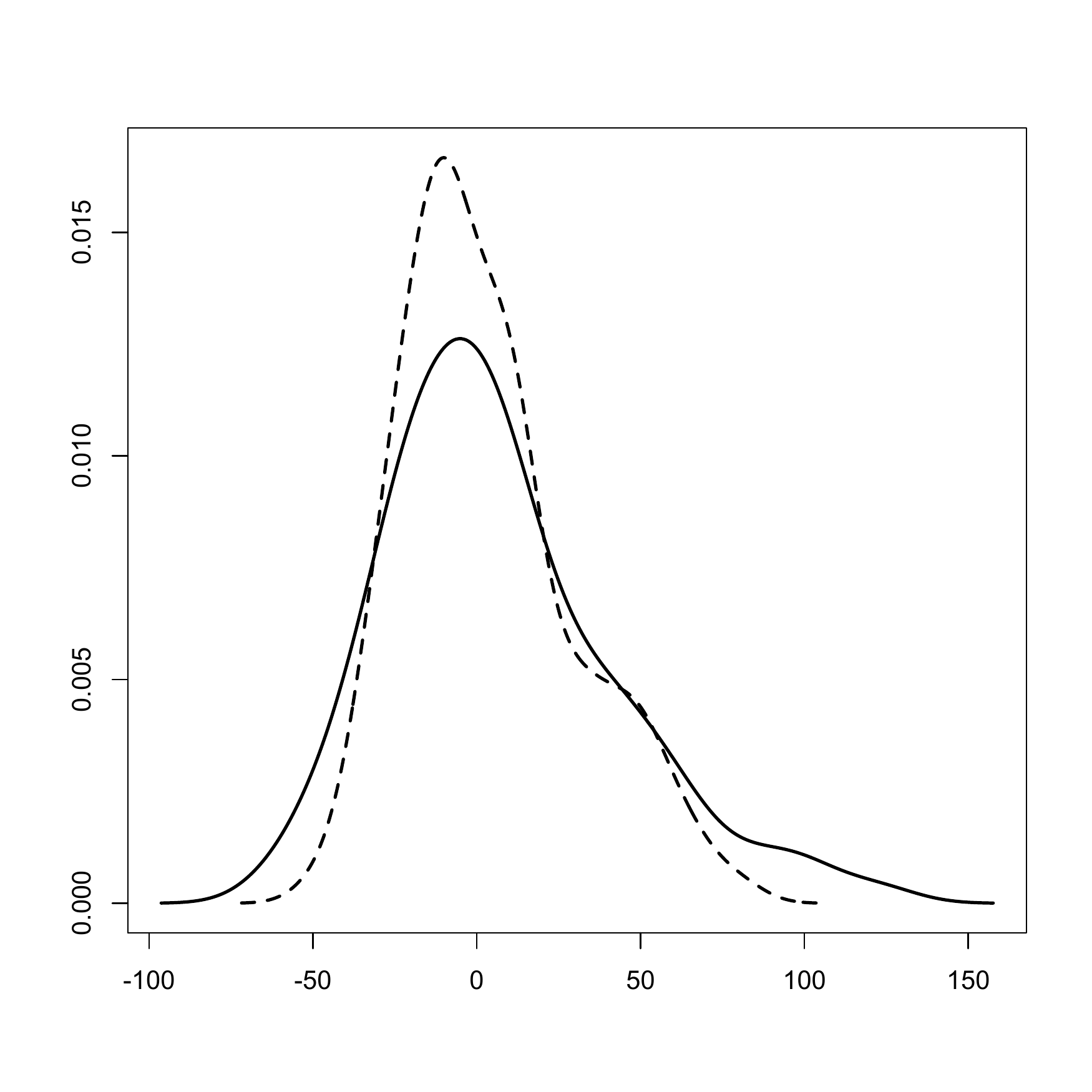} 
\end{center}
\caption{\textit{Fowl bones data set}: densities of the relative errors on the posterior probabilities for the $107$ most probable graphs. 
\textit{add and delete} kernel in solid line and data-driven kernel in dashed line.}
\label{fig:density-biais-fowlbones}
\end{figure*}

These values are compared to the theoretical ones $p(\G_{int}| \bY;\tau^*,r^*)$. In Figure \ref{fig:density-biais-fowlbones},
we plot the estimated densities of the quantities relative errors 
$$\frac{\widehat{\pi}_1 (\G_{int} |   \bY;\tau^*,r^*)  -p(\G_{int} |   \bY;\tau^*,r^*)}{p(\G_{int} |   \bY;\tau^*,r^*)} \times 100$$ in solid line, and
$$\frac{\widehat{\pi}_2 (\G_{int} |   \bY;\tau^*,r^*)  -p(\G_{int} |   \bY;\tau^*,r^*)}{p(\G_{int} |   \bY;\tau^*,r^*)} \times 100$$ in dashed line. 

\begin{figure*}[htb]
\begin{center}
\includegraphics[width=0.6\textwidth]{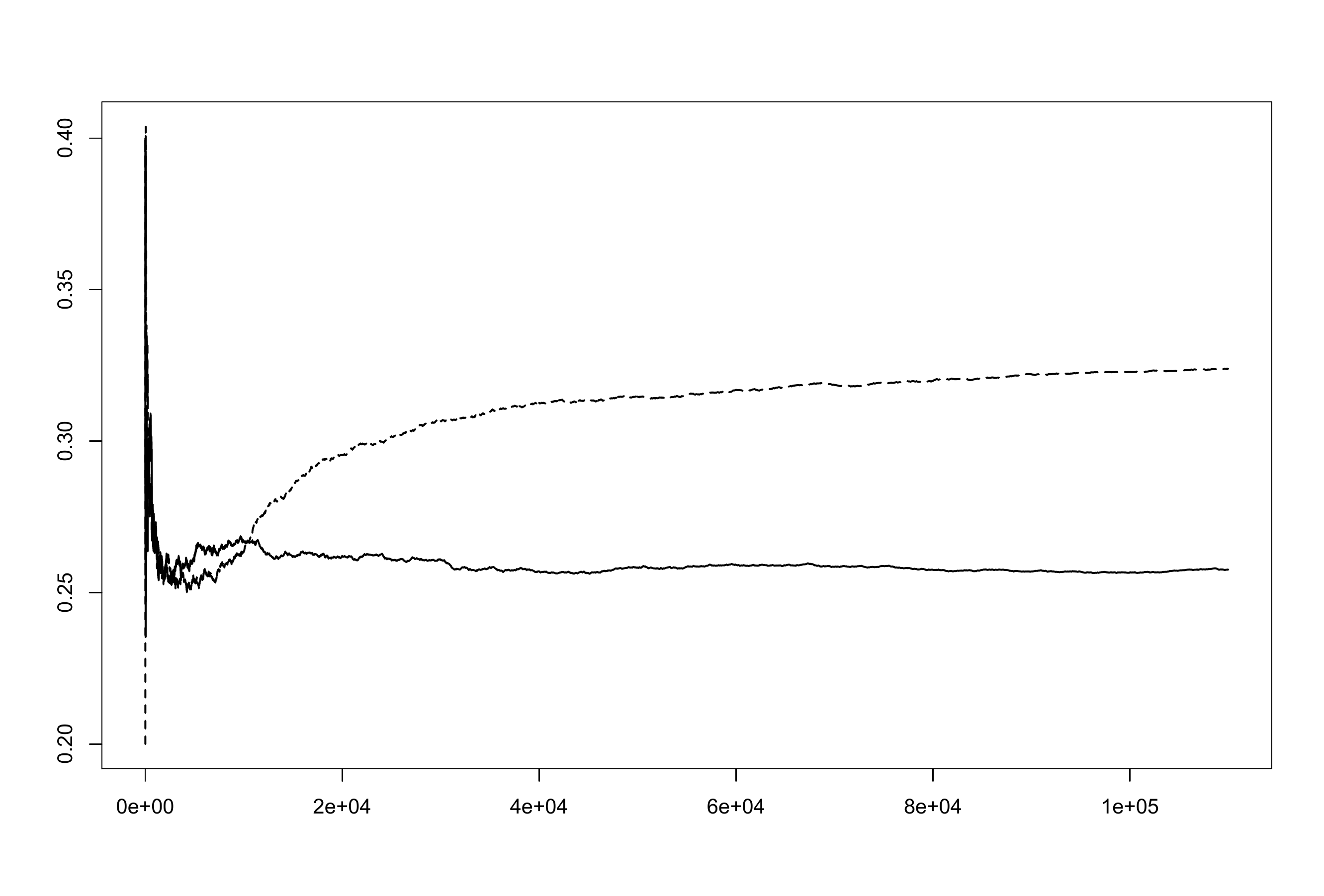} 
\end{center}
\caption{\textit{Fowl bones data set}: evolution of the acceptance ratio for the \textit{add and delete} Markov chains (solid line)
and the data driven Markov chains ($-\:$ dashed line). }
\label{fig:fowl_taux_accept}
\end{figure*}

We note that the density corresponding to the errors involved by the data-driven kernel is more concentrate around the value $0$.
The large errors in the \textit{add an delete} density are due to the graphs with small probabilities. Thus, the new kernel explores
more efficiently the posterior distribution. The acceptance rate is higher for the data-driven chain  
(see Figure \ref{fig:fowl_taux_accept}).

\section{Conclusion and discussion}

An empirical Bayes strategy estimating prior hy\-per-parameters in a Gaussian graphical model using a SAEM-MCMC algorithm is introduced. \\
That proposal does not depend on any calibrating parameters and can be viewed as a default option for decomposable graphical model determination.
Some empirical studies show the relevance of the proposed approach and the good properties of the introduced algorithms.

However, \cite{scott:berger:2010} has recently found considerable differences between fully Ba\-yes and empirical Bayes strategies in the context
of variable selection. It would be very interesting to investigate, both from theoretical and practical perspectives, on such a discrepancy in the
case of decomposable graphical model selection.

\section*{Acknowledgments}

The authors are grateful to Marc Lavielle for very helpful discussions.
The authors wish to thank the Associate Editor and two reviewers
whose suggestions were very helpful in improving the presentation of this work.
This work has been supported by the Agence Nationale de la Recherche (ANR, 212, rue de Bercy 75012 Paris)
through the 2009-2012 project Big'MC.

\bibliographystyle{apalike}  
\bibliography{DM10}

\end{document}